\documentclass[aps,showpacs,superscriptadress,preprint]{revtex4}
\usepackage{graphicx}
\usepackage{amsmath}
\usepackage{amssymb}
\usepackage{mathrsfs}
\usepackage{stmaryrd}
\usepackage{amsthm}
\usepackage{hyperref}
\usepackage{multirow}
\hypersetup{hypertex=true,
	colorlinks=true,
	linkcolor=red,
	anchorcolor=green,
	citecolor=blue}
\usepackage{caption}
\usepackage{tikz,xcolor,hyperref}

\definecolor{lime}{HTML}{A6CE39}
\DeclareRobustCommand{\orcidicon}{
\begin{tikzpicture}
\draw[lime, fill=lime] (0,0)
circle[radius=0.16]
node[white]{{\fontfamily{qag}\selectfont \tiny \.{I}D}}; 
\end{tikzpicture}
\hspace{-3mm}
}
\foreach \x in {A, ..., Z}{%
\expandafter\xdef\csname orcid\x\endcsname{\noexpand\href{https://orcid.org/\csname orcidauthor\x\endcsname}{\noexpand\orcidicon}}
}

\begin{document}
\title{Standard constraints on the shadow radius of black holes in $d$-dimensional space-time: The case of regular Einstein-Gauss-Bonnet solutions}

\author{
Zening Yan$^{1}$\footnote{Corresponding author: z.n.yan.bhtpr@gmail.com or znyan21@m.fudan.edu.cn}\hspace{-2mm}\orcidA{}
} 

\affiliation{
\small 1. Center for Astronomy and Astrophysics \& Center for Field Theory and Particle Physics \& Department of Physics, Fudan University, No.2005 Songhu Road, Yangpu District, Shanghai 200438, China\\
}


\begin{abstract}
In this work, we propose a $d$-dimensional constraint formula for the black hole shadow radius, providing a standardized framework for using observational confidence intervals to constrain the corresponding characteristic parameters in $d$-dimensional black hole solutions. 
This approach can correct the misapplication of shadow radius constraints in some previous studies.
Subsequently, we selected two regular Einstein-Gauss-Bonnet (EGB) black hole solutions as a representative example to test the $d$-dimensional shadow constraint formula.
Our analysis reveals that the characteristic parameters within these $d$-dimensional black hole solutions maintain physically meaningful bounded domains under shadow radius constraints.
We further propose that this $d$-dimensional shadow constraint formula is universal and can be applied to constrain the characteristic parameters in other high-dimensional spherically symmetric black hole solutions.
\end{abstract}


\maketitle

\tableofcontents

\section{Introduction}
The black hole shadow, defined as a photon-deficient region projected on the celestial sphere through extreme gravitational lensing, manifests as a central dark disk bounded by a luminous photon ring.
This resurgence of research interest stems from recent revolutionary advances in observational astrophysics, particularly the landmark achievement of horizon-scale imaging highlighted by the Event Horizon Telescope's $\mathrm{M87}^{\star}$ revelation.

The theoretical foundations for detecting black hole shadow imaging were established in 2000 through the pioneering work of Falcke et al \cite{Falcke:1999pj}.
These advancements culminated in 2019 when the Event Horizon Telescope (EHT) collaboration institution achieved a historic milestone by capturing the first resolved images of the supermassive black hole $\mathrm{M87}^{\star}$ \cite{EventHorizonTelescope:2019dse}.
Building on this success, the EHT further revealed in 2022 high-resolution visuals of the supermassive black hole Sagittarius $\mathrm{A}^{\star}$ ($\mathrm{Sgr~A}^{\star}$), located at the center of the Milky Way \cite{EventHorizonTelescope:2022wkp}.
Afterwards, the EHT successively unveiled polarized views of $\mathrm{M87}^{\star}$ \cite{EventHorizonTelescope:2021bee} and $\mathrm{Sgr~A}^{\star}$ \cite{EventHorizonTelescope:2024rju} pertaining to the magnetic field surrounding them.
In April 2023, an international team of scientists led by Ru-Sen Lu released a panoramic image capturing both the black hole shadow and its jet in the Messier 87 (M87) galaxy.
This was achieved by combining observations from the Global Millimeter VLBI Array (GMVA), the Atacama Large Millimeter/submillimeter Array (ALMA) and the Greenland Telescope (GLT) \cite{Lu:2023bbn}.
Concurrently, Lia Medeiros and collaborators introduced a new technique called Principal-component Interferometric Modeling (PRIMO), which enhances EHT image data to reconstruct a sharper shadow image of the black hole \cite{Medeiros:2023pns}.
In August 2024, the EHT achieved the highest diffraction-limited angular resolution ever attained on Earth by operating at a frequency of 345 $\text{GHz}$.
This advancement enabled the capture of significantly clearer shadow images of black holes.
It is anticipated that the new results will enhance the clarity of the image by 50\%.

Furthermore, the EHT collaboration has established observationally constrained confidence intervals for the shadow radii of black holes $\mathrm{M87}^{\star}$ and $\mathrm{Sgr~A}^{\star}$.
These measurements provide critical empirical benchmarks for testing modified gravity theories, enabling quantitative constraints on space-time metric parameters across multiple theoretical frameworks --- including the Reissner-Nordstr\"{o}m (RN), Kerr, Bardeen, Hayward, Frolov, Janis-Newman-Winicour (JNW), Kazakov-Solodhukin (KS), and Einstein-Maxwell-dilaton (EMd-1) \cite{EventHorizonTelescope:2021dqv, EventHorizonTelescope:2020qrl, EventHorizonTelescope:2022xqj}.
Subsequently, the characteristic parameters of many other black holes have been further constrained \cite{Khodadi:2022pqh, Vagnozzi:2022moj, Yan:2023pxj, Uniyal:2022vdu, Pantig:2022qak, Yan:2024rsx}.
However, we have observed that some studies incorrectly apply the constraint formula for the four-dimensional black hole shadow radius derived from Schwarzschild modeling to certain $d$-dimensional black hole solutions, which is obviously inappropriate.
Therefore, this paper will provide a constraint formula for the shadow radius applicable to $d$-dimensional black holes.

This paper is organized as follows:
In Section $\mathrm{\uppercase\expandafter{\romannumeral2}}$, we propose a constraint formula for the shadow radius that is applicable to $d$-dimensional black hole space-times.
In Section $\mathrm{\uppercase\expandafter{\romannumeral3}}$, we introduced two regular EGB black hole solutions and operated on them using the $d$-dimensional shadow constraint formula.
In Section $\mathrm{\uppercase\expandafter{\romannumeral4}}$, we determined the reasonable ranges for the characteristic parameters $(\alpha, q)$ in the two black hole solutions, thereby confirming the validity and accuracy of the $d$-dimensional shadow constraint formula.
In Section $\mathrm{\uppercase\expandafter{\romannumeral5}}$, we provide a brief summary of the entire article.

Throughout this work, we employ units in which the speed of light is set to unity ($c=1$).

\section{Constraints of high-dimensional shadows}
The metric of a static spherically symmetric black hole in $d$-dimensional space-time can be written as
\begin{equation}\label{eq:metric}
\mathrm{d}s^2 = -h(r)\mathrm{d}t^2 + \frac{1}{f(r)}\mathrm{d}r^2 + r^2 \mathrm{d}\Omega_{d-2}^2,
\end{equation}
where $\mathrm{d}\Omega_{d-2}^2$ is the line element on the $(d-2)$-dimensional unit sphere.
The general formula for the shadow radius of a $d$-dimensional spherically symmetric black hole has been established as
\begin{equation}\label{eq:highdimrs}
R_{s} = \sqrt{ \frac{\left(r_p\right)^3}{h\left(r_p\right) \big[r_p f^{\prime}\left(r_p\right)+ 2 f\left(r_p\right)\big]} \Bigg\{ f^{\prime}\left(r_p\right) - f\left(r_p\right) \bigg[ \frac{h^{\prime}\left(r_p\right)}{h\left(r_p\right)} - \frac{4}{r_p} \bigg] \Bigg\} },
\end{equation}
and its detailed derivation process is provided in paper \cite{Nozari:2023flq}, which will not be elaborated on further here.
In this section, we will focus on demonstrating a correct constraint equation when utilizing observational data from EHT to constrain certain high-dimensional black holes.

Previously, the stellar-dynamics measurements of supermassive black holes were used to produce a posterior distribution function of the angular gravitational radius $\theta_{\mathrm{g}}$, which was determined by the EHT in 2017 through observations of the $\mathrm{M87}^{\star}$ and $\mathrm{Sgr~A}^{\star}$ black holes.
The EHT team used the prior information on the angular gravitational radius $\theta_{\mathrm{g}}$ of the $\mathrm{M87}^{\star}$ and $\mathrm{Sgr~A}^{\star}$ black holes to calculate the predicted size of their shadows. 
Afterwards, the posterior of $\theta_{\mathrm{g}}$ was analyzed by comparing its prior results with the size inferred from the EHT images and visibility-domain model fitting.

The angular gravitational radius $\theta_{\mathrm{g}}$ is defined as
\begin{equation}\label{eq:agr}
\theta_{\mathrm{g}} = \frac{\mathrm{G} M}{c^2 D} = \frac{\mathrm{G} M}{D},
\end{equation}
where $M$ is the mass of the black hole and $D$ is the distance between the black hole and the observer.
Next, the fractional deviation of the angular gravitational radius is introduced as follows:
\begin{equation}\label{eq:agrfd}
\delta \equiv \frac{\theta_{\mathrm{dyn}}}{\theta_{\mathrm{g}}} - 1,
\end{equation}
where $\theta_{\mathrm{g}}$ and $\theta_{\mathrm{dyn}}$ are used to represent the real measurement and the inference of the stellar-dynamics of the angular gravitational radius, respectively.
Thus, $\theta_{\mathrm{dyn}}$ can be written as
\begin{equation}
\theta_{\mathrm{dyn}} = \frac{\mathrm{G} M_{\mathrm{dyn}}}{D_{\mathrm{dyn}}}.
\end{equation}
In addition, the angular shadow radius can be defined as
\begin{equation}\label{eq:asr}
\theta_{\mathrm{sh}} = \frac{r_{\mathrm{sh}}}{D},
\end{equation}
then, with the known value of the shadow radius $r_{\mathrm{sh}}$, the predicted value of the angular shadow radius $\theta_{\mathrm{sh}}$ can be determined using $\theta_{\mathrm{dyn}}$.

Since the stellar-dynamics measurements are only sensitive to the monopole of the metric (i.e., the mass) and spin-dependent effects are negligible at the distances involved in that analysis, it is reasonable to model the $\mathrm{M87}^{\star}$ and $\mathrm{Sgr~A}^{\star}$ using the Schwarzschild solution.
The EHT has given several limits on the fractional deviations from different observation instruments \cite{EventHorizonTelescope:2019ggy, EventHorizonTelescope:2022xqj}, as shown in Table \ref{tb1:deltadata}.
And then, reasonable deviations in the shadow radius size can be obtained.
Since these deviations can still be consistent with the imaging data, an effective constraint range can be imposed on certain underlying parameters in the black hole metric \cite{EventHorizonTelescope:2021dqv, EventHorizonTelescope:2020qrl}.
\begin{table}[tbh]\centering
\caption{Reasonable boundaries of fractional deviation $\delta$ are provided from different observation instruments.
\vspace{0.2cm}} \label{tb1:deltadata}
\begin{tabular*}{12cm}{*{3}{c @{\extracolsep\fill}}}
\hline
\textbf{Black Hole} & \textbf{Observation Instrument} & $\delta$ \\
\hline
$\mathrm{M87}^{\star}$ & \text{EHT} & $-0.01_{-0.17}^{+0.17}$ \\
\multirow{2}{*}{$\mathrm{Sgr~A}^{\star}$} & \text{VLTI} & $-0.08_{-0.09}^{+0.09}$ \\
                                          & \text{Keck} & $-0.04_{-0.10}^{+0.09}$ \\ 
\hline
\end{tabular*}
\end{table}

When we choose the Schwarzschild solution in four-dimensional space-time as the predicted value of the shadow radius $r_{\mathrm{sh}}$, it can be expressed as
\begin{equation}
r_{\mathrm{sh}} = 3\sqrt{3} G M_{\mathrm{dyn}},
\end{equation}
where $G$ is the four-dimensional gravitational constant, and then $\theta_{\mathrm{sh}}$ can be determined as
\begin{equation}
\theta_{\mathrm{sh}} = \frac{r_{\mathrm{sh}}}{D_{\mathrm{dyn}}} = 3\sqrt{3} \left( \frac{G M_{\mathrm{dyn}}}{D_{\mathrm{dyn}}} \right),
\end{equation}
therefore, $\theta_{\mathrm{sh}} = 3\sqrt{3} \theta_{\mathrm{dyn}}$, and substituting it into Eq. \eqref{eq:agrfd}, we can get
\begin{equation}\label{eq:thetashdelta}
\theta_{\mathrm{sh}} = 3\sqrt{3} \left(\delta + 1\right) \theta_{\mathrm{g}}.
\end{equation}
Utilizing Eqs. \eqref{eq:agr} and \eqref{eq:thetashdelta}, it is natural to get the following relation:
\begin{equation}
\theta_{\mathrm{sh}} = \frac{r_{\mathrm{sh}}}{D} = \frac{r_{\mathrm{sh}}}{\left( \frac{G M}{\theta_{\mathrm{g}}} \right)} = 3\sqrt{3} \left(\delta + 1\right) \theta_{\mathrm{g}},
\end{equation}
so the shadow radius is denoted by
\begin{equation}\label{eq:4dimrange}
r_{\mathrm{sh}} = 3\sqrt{3} \left(\delta + 1\right) G M.
\end{equation}
In this way, we have constructed a reasonable constraint on the shadow radius of a four-dimensional black hole using the Schwarzschild metric.

However, it is incorrect for certain studies to apply Eq. \eqref{eq:4dimrange} to $d$-dimensional black hole solutions \cite{Nozari:2024jiz, Nozari:2023flq}.
Firstly, Eq. \eqref{eq:4dimrange} is derived by modeling the theoretical value of the shadow radius of the four-dimensional Schwarzschild black hole.
Consequently, it is evident that the shadow radius of a high-dimensional black hole cannot logically be constrained within this four-dimensional framework.
Secondly, if Eq. \eqref{eq:4dimrange} is forcibly applied to constrain the characteristic parameters in high-dimensional black hole solutions, unphysical scenarios may arise. 
For instance, certain characteristic parameters could erroneously appear as constants or fixed functions as the dimension grows, which is clearly impossible.

To ensure the validity of the constraint range is valid in $d$-dimensional black hole solutions, we derive the shadow radius constraint equation specifically for black holes in $d$-dimensional space-time.
Here, we will use the shadow radius of the Schwarzschild-Tangherlini metric as the predicted value for modeling, and its radius formula can be expressed as
\begin{equation}
r_{\mathrm{sh}} = \sqrt{ \frac{d-1}{d-3} \left[ \frac{16 \pi \mathscr{G}M_{\mathrm{dyn}} (d-1)}{2 (d-2) \Sigma_{(d-2)}} \right]^{\frac{2}{d-3}} },
\end{equation}
and then,
\begin{equation}\label{eq:dthsh}
\begin{aligned}
\theta_{\mathrm{sh}} &= \frac{r_{\mathrm{sh}}}{D_{\mathrm{dyn}}} \\
&= \left\{ \left(\frac{d-1}{d-3}\right)^{\frac{1}{2}} \left[ \frac{16 \pi (d-1)}{2 (d-2) \Sigma_{(d-2)}} \right]^{\frac{1}{d-3}} \cdot \left(\mathscr{G} M_{\mathrm{dyn}}\right)^{\frac{1}{d-3}} \right\} \cdot \frac{1}{D_{\mathrm{dyn}}}.
\end{aligned}
\end{equation}
Next, we define the function $\mathfrak{C}(d)$ as
\begin{equation}
\mathfrak{C}(d) \equiv \left(\frac{d-1}{d-3}\right)^{\frac{1}{2}} \left[ \frac{16 \pi (d-1)}{2 (d-2) \Sigma_{(d-2)}} \right]^{\frac{1}{d-3}},
\end{equation}
so that Eq. \eqref{eq:dthsh} can be rewritten as
\begin{equation}\label{eq:dthsh2}
\begin{aligned}
\theta_{\mathrm{sh}} &= \mathfrak{C}(d) \cdot \left(\mathscr{G} M_{\mathrm{dyn}}\right)^{\frac{1}{d-3}} \cdot \frac{1}{D_{\mathrm{dyn}}} \\
&= \mathfrak{C}(d) \cdot \left(\mathscr{G} M_{\mathrm{dyn}}\right)^{\frac{1}{d-3} - 1} \cdot \left(\frac{\mathscr{G} M_{\mathrm{dyn}}}{D_{\mathrm{dyn}}}\right) \\
&= \mathfrak{C}(d) \cdot \left(\mathscr{G} M_{\mathrm{dyn}}\right)^{\frac{1}{d-3} - 1} \cdot \theta_{\mathrm{dyn}}. \\
\end{aligned}
\end{equation}
Substituting Eq. \eqref{eq:dthsh2} into Eq. \eqref{eq:agrfd}, we can get
\begin{equation}
\delta = \frac{\left[ \frac{\theta_{\mathrm{sh}}}{\mathfrak{C}(d) \cdot \left(\mathscr{G} M_{\mathrm{dyn}}\right)^{\frac{1}{d-3} - 1}} \right]}{\theta_{\mathrm{g}}} - 1,
\end{equation}
that is,
\begin{equation}\label{eq:dthetashdelta}
\theta_{\mathrm{sh}} = \left[ \mathfrak{C}(d) \cdot \left(\mathscr{G} M_{\mathrm{dyn}}\right)^{\frac{1}{d-3} - 1} \right] \left(\delta + 1\right) \theta_{\mathrm{g}}.
\end{equation}
Combining Eqs. \eqref{eq:agr} and \eqref{eq:dthetashdelta}, we have
\begin{equation}
\theta_{\mathrm{sh}} = \frac{r_{\mathrm{sh}}}{D} = \frac{r_{\mathrm{sh}}}{\left( \frac{\mathscr{G} M}{\theta_{\mathrm{g}}} \right)} = \left[ \mathfrak{C}(d) \cdot \left(\mathscr{G} M_{\mathrm{dyn}}\right)^{\frac{1}{d-3} - 1} \right] \left(\delta + 1\right) \theta_{\mathrm{g}},
\end{equation}
therefore, the shadow radius is given by
\begin{equation}
r_{\mathrm{sh}} = \left[ \mathfrak{C}(d) \cdot \left(\mathscr{G}\right)^{\frac{1}{d-3}} \cdot \left(M_{\mathrm{dyn}}\right)^{\frac{1}{d-3} - 1} \right] \left(\delta + 1\right) M.
\end{equation}
It should be noted here that $\mathscr{G}$ and  $M_{\mathrm{dyn}}$ depend on the dimension $d$.
We can refer to the extra-dimensional compactification volume $\mathscr{V}_{(d-4)}$ to relate the high-dimensional gravitational constant $\mathscr{G}$ to the four-dimensional gravitational constant $G$, and the relationship is defined in terms of 
\begin{equation}
G = \frac{\mathscr{G}}{\mathscr{V}_{(d-4)}}.
\end{equation}
If the extra dimension is a flat torus with $d$ dimensions, then $\mathscr{V}_{(d-4)}=\left(2\pi\mathscr{R}\right)^{d-4}$.
Thus, the high-dimensional gravitational constant $\mathscr{G}$ can be expressed as
\begin{equation}
\mathscr{G} = G \left(2\pi\mathscr{R}\right)^{d-4},
\end{equation}
where $\mathscr{R}$ represents the extra-dimensional compactification radius for each dimension $d$.
Accordingly, the shadow radius $r_{\mathrm{sh}}$ can be rewritten as
\begin{equation}
r_{\mathrm{sh}} = \left[ \mathfrak{C}(d) \cdot \left(G\right)^{\frac{1}{d-3}} \cdot \left(\frac{M_{\mathrm{dyn}}}{2\pi\mathscr{R}}\right)^{\frac{1}{d-3} - 1} \right] \left(\delta + 1\right) M.
\end{equation}
If we consider stipulating $M_{\mathrm{dyn}}=2\pi\mathscr{R}$ for each dimension $d$, the above formula can be simplified to
\begin{equation}\label{eq:highdimrange}
r_{\mathrm{sh}} = \left[ \mathfrak{C}(d) \cdot \left(G\right)^{\frac{1}{d-3}} \right] \left(\delta + 1\right) M.
\end{equation}

Since the Gauss-Bonnet term serves merely as a higher-order curvature correction, the four-dimensional gravitational constant $G$ after compactification can still be expressed as the ratio of the high-dimensional gravitational constant $\mathscr{G}$ to the extra-dimensional compactification volume $\mathscr{V}_{(d-4)}$.
Therefore, the $\mathscr{G} M$ term in the metric can also be reformulated as $G \left(2\pi\mathscr{R}\right)^{d-3}$ in compliance with the stipulation $M=2\pi\mathscr{R}$.
In actual numerical calculations, in order to obtain the value of $R_{s}$ (i.e., Eq. \eqref{eq:highdimrs}) corresponding to the metric, we will further specify $M=2\pi\mathscr{R}=1$ for convenience.
Note that this specification implies a fixed value selected when calculating a certain dimension $d$, rather than indicating that it holds for different dimensions.
When we do not consider calculations between different dimensions, the above specification is allowed.

It can be observed that different dimensions correspond to different constraint domains $r_{\mathrm{sh}}$ of shadow radius in $d$-dimensional space-time.
Additionally, an intriguing phenomenon is that for the $d$-dimensional constraint equation, the contribution of ${G}$ to $r_{\mathrm{sh}}$ is affected by the dimension $d$ when ${G}\neq1$.

Therefore, for the sake of standardization of shadow radius constraints, when we systematically constrain the characteristic parameters in $d$-dimensional black hole solutions, Eq. \eqref{eq:highdimrange} should be used.

\section{Regular black holes in Einstein-Gauss-Bonnet theory}
The action of the EGB theory of gravity minimally coupled to a non-linear electromagnetic field can be expressed as
\begin{equation}\label{eq:action}
\mathcal{S}= \frac{1}{16 \pi \mathscr{G}} \int_{\mathcal{M}} \mathrm{d}^d x \sqrt{-g}\Big[\mathcal{R}+\alpha\left(\mathcal{R}^2-4 \mathcal{R}_{\mu \nu} \mathcal{R}^{\mu \nu}+\mathcal{R}_{\mu \nu \gamma \delta} \mathcal{R}^{\mu \nu \gamma \delta}\right)+\mathcal{L}(F)\Big],
\end{equation}
where $\mathscr{G}$ and $\alpha$ are the $d$-dimensional gravitational constant and Gauss-Bonnet coupling coefficient, respectively.
And the non-linear electromagnetic field is described by $\mathcal{L}(F)$.

When considering a static spherically symmetric metric \eqref{eq:metric} that is asymptotically flat in $d$-dimensional space-time, we will present two exact regular solutions in EGB gravity: the Bardeen-EGB black hole \cite{Kumar:2018vsm} and the Hayward-EGB black hole \cite{Ghosh:2020tgy}, which are expressed as
\begin{equation}\label{eq:b-egb}
h(r)=f(r)=1+\frac{r^2}{2 \tilde{\alpha}}\left(1 \mp \sqrt{1+\frac{4 \tilde{\alpha} m}{\left(r^{d-2}+q^{d-2}\right)^{\frac{d-1}{d-2}}}}\right),
\end{equation}
and
\begin{equation}\label{eq:h-egb}
h(r)=f(r)=1+\frac{r^2}{2 \tilde{\alpha}}\left(1 \mp \sqrt{1+\frac{4 \tilde{\alpha} m}{r^{d-1}+q^{d-1}}}\right),
\end{equation}
where $\tilde{\alpha}=(d-3)(d-4)\alpha$, the symbol $q$ represents the magnetic charge and the mass term $m$ is a constant of integration and is related to the Arnowitt-Deser-Misner (ADM) mass $M$ of the black hole via
\begin{equation}
m=\frac{16 \pi \mathscr{G} M}{(d-2) \Sigma_{(d-2)}}.
\end{equation}
The symbol $\Sigma_{(d-2)}$ represents the volume of the $(d-2)$-dimensional unit sphere, which can be written as $\Sigma_{(d-2)}=\frac{2 \pi^{(d-1)/2}}{\Gamma\left(\frac{d-1}{2}\right)}$.
In addition, the branch ``$+$'' is unstable for small perturbations, so only the branch ``$-$'' is physical.

When we choose zero magnetic charge (i.e., $q=0$) and take the limit $\alpha \rightarrow 0$, Eqs. \eqref{eq:b-egb} and \eqref{eq:h-egb} will be reduced to the classical Schwarzschild-Tangherlini solution:
\begin{equation}
f(r)=1-\frac{16 \pi \mathscr{G} M}{(d-2) \Sigma_{(d-2)} r^{d-3}}.
\end{equation}
When restricted to the case of $q=0$, Eqs. \eqref{eq:b-egb} and \eqref{eq:h-egb} reduce to the $d$-dimensional EGB black hole solution \cite{Cai:2001dz, Cho:2002hq}, with the metric given by:
\begin{equation}\label{eq:egbmetric}
f(r)=1+\frac{r^2}{2 \tilde{\alpha}}\left(1 \mp \sqrt{1+\frac{4 \tilde{\alpha} m}{r^{d-1}}}\right).
\end{equation}

\section{Numerical results}
\begin{figure}[htbp]
\centering
\includegraphics[width=1\textwidth]{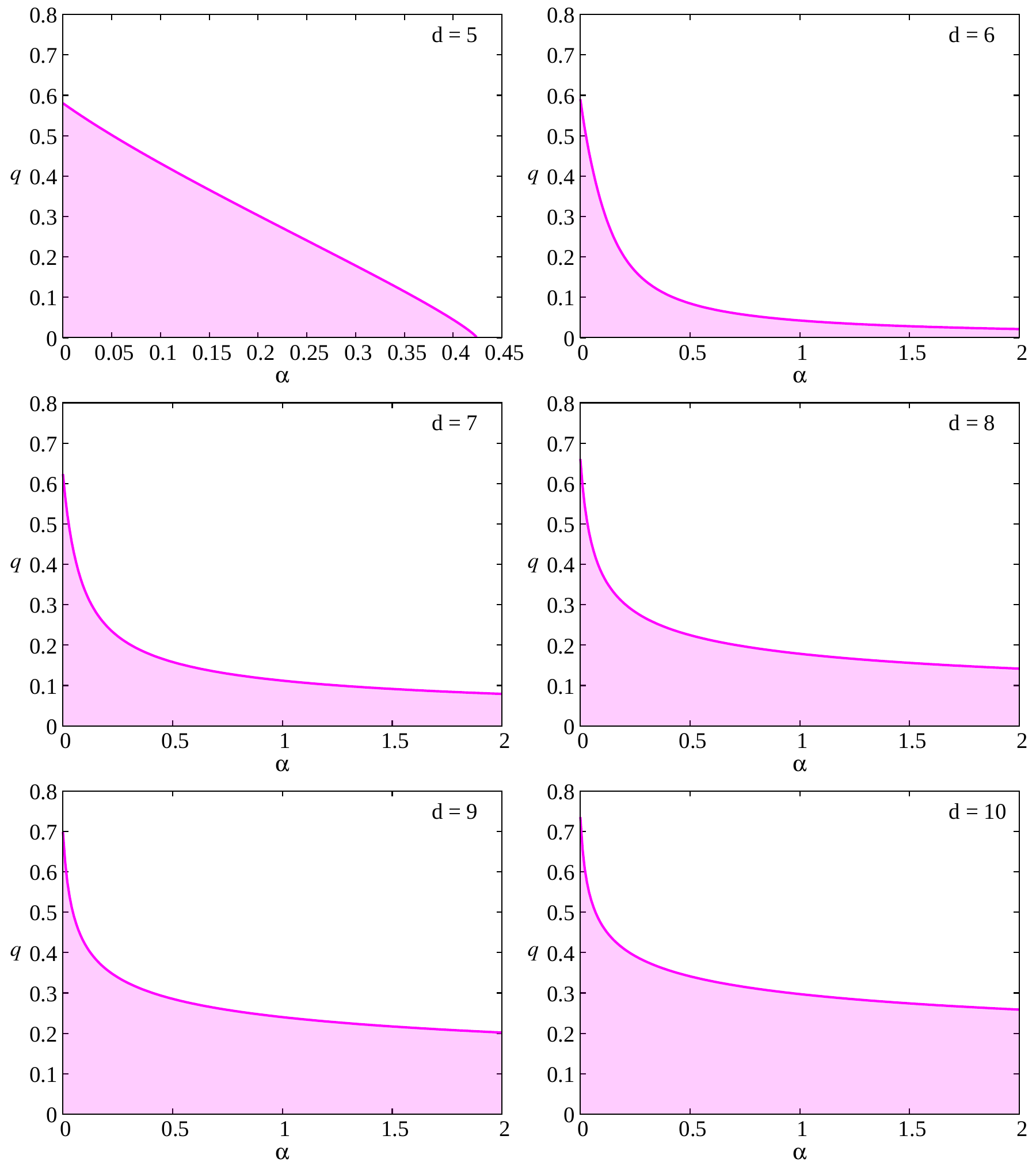}
\caption{The colored area (\textit{magenta}) indicates the valid range of parameters $(\alpha, q)$ when the Bardeen-EGB black hole ensures the existence of the event horizon. The parameters $M=1$ and $\mathscr{G}=1$ are selected.}
\label{FIG1}
\end{figure}

\begin{figure}[htbp]
\centering
\includegraphics[width=1\textwidth]{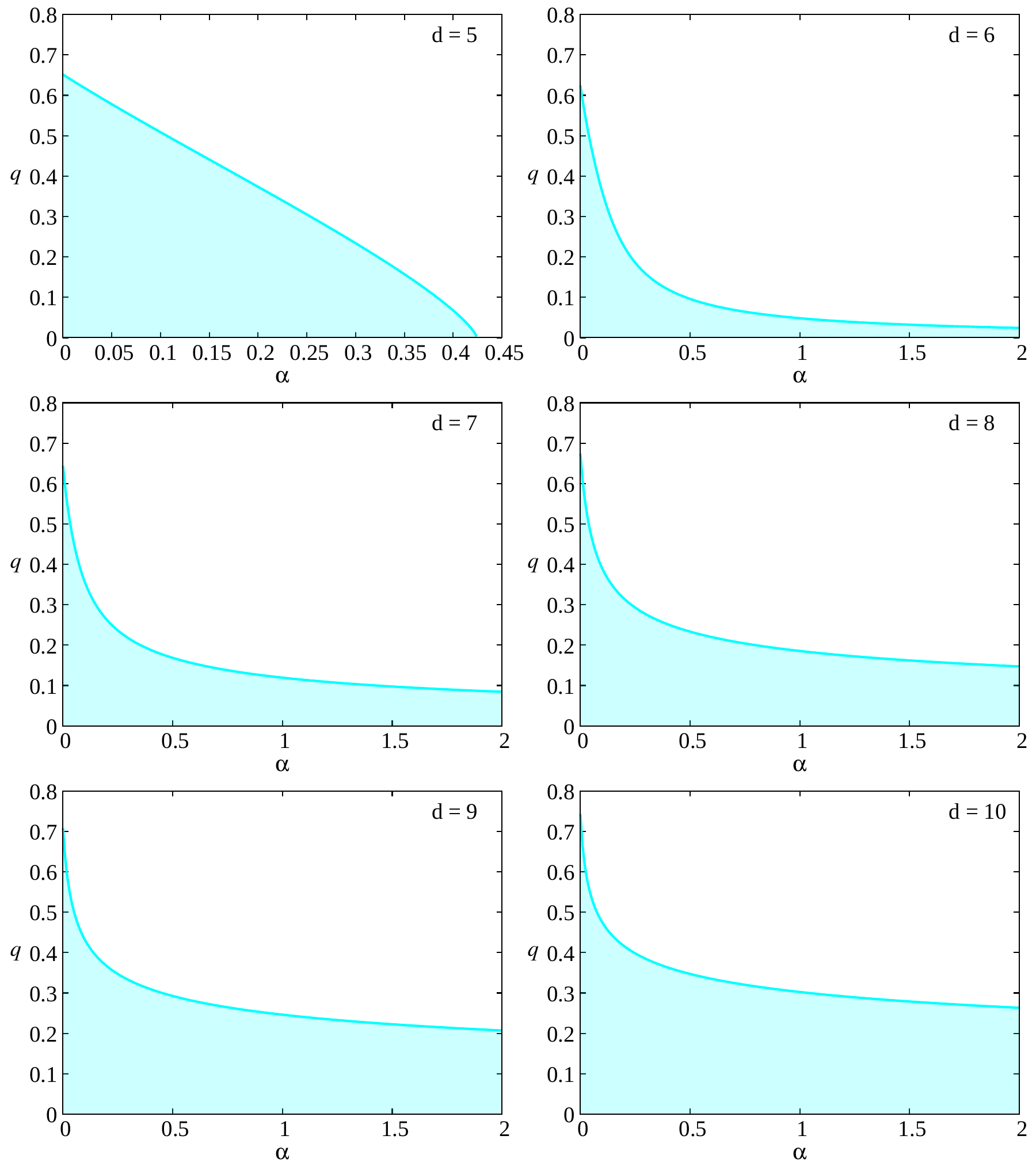}
\caption{The colored area (\textit{cyan}) indicates the valid range of parameters $(\alpha, q)$ when the Hayward-EGB black hole ensures the existence of the event horizon. The parameters $M=1$ and $\mathscr{G}=1$ are selected.}
\label{FIG2}
\end{figure}

Firstly, in order to ensure that the black hole solution has at least one event horizon, the values of the parameters $\alpha$ and $q$ will be confined to a valid region, as shown in Figures \ref{FIG1} and \ref{FIG2}.

Notably, the parameter ranges for five-dimensional space-time exhibit distinct characteristics compared to higher-dimensional cases in both black hole solutions. 
Specifically, when the space-time dimension is $d=5$, the parameter $\alpha$ is confined to a bounded interval ($0 < \alpha \lesssim 0.42$) determined by the existence conditions of horizons.

\begin{figure}[htbp]
\centering
\includegraphics[width=1\textwidth]{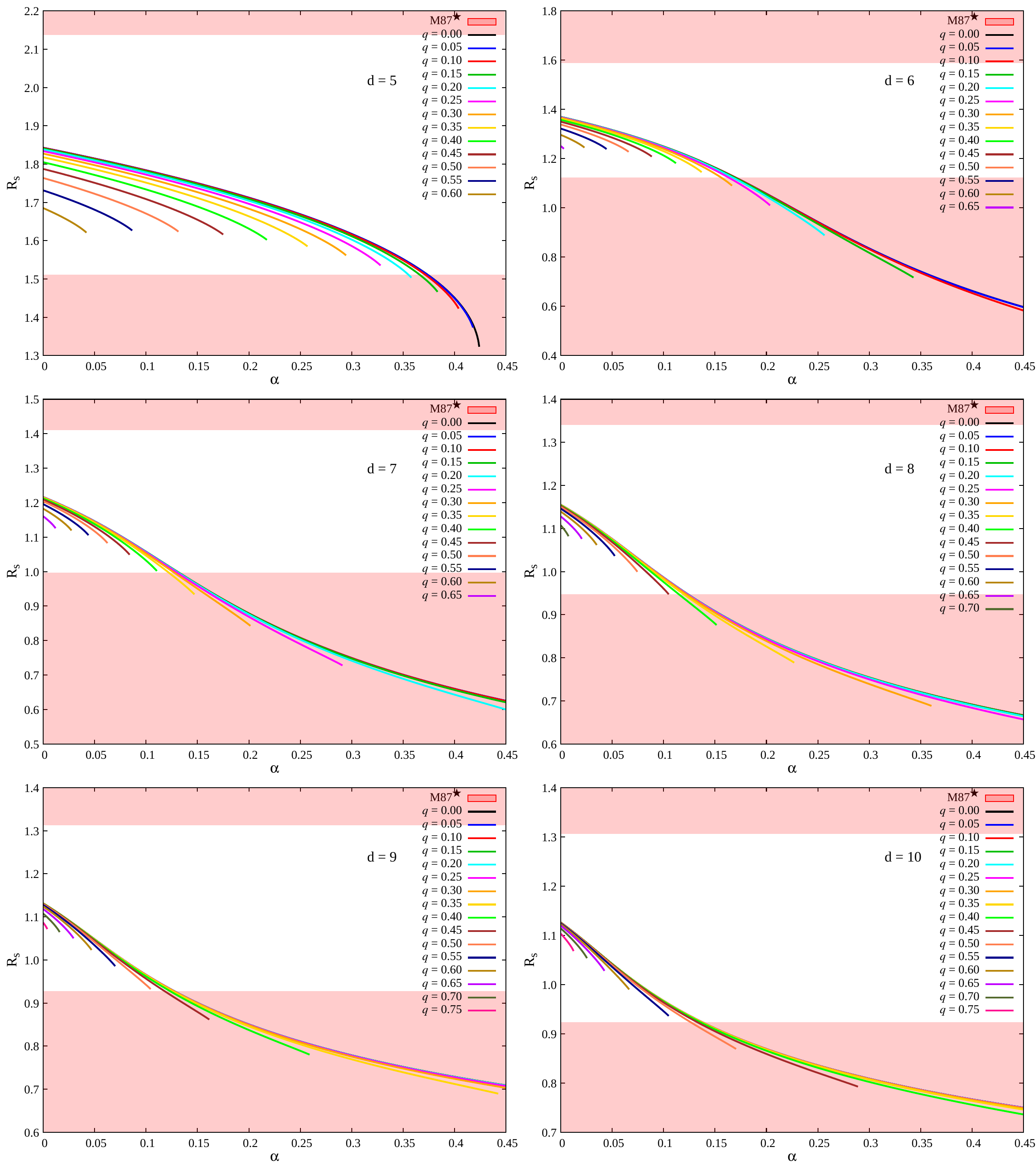}
\caption{The relevant parameters of the Bardeen-EGB black hole are constrained through the $d$-dimensional shadow constraint formula \eqref{eq:highdimrange}. The unfilled color region corresponds to the effective range $r_{\mathrm{sh}}$ of the shadow radius from the $\mathrm{M87}^{\star}$ data. The parameters $M=1$ and ${G}=1$ are selected.}
\label{FIG3}
\end{figure}

\begin{figure}[htbp]
\centering
\includegraphics[width=1\textwidth]{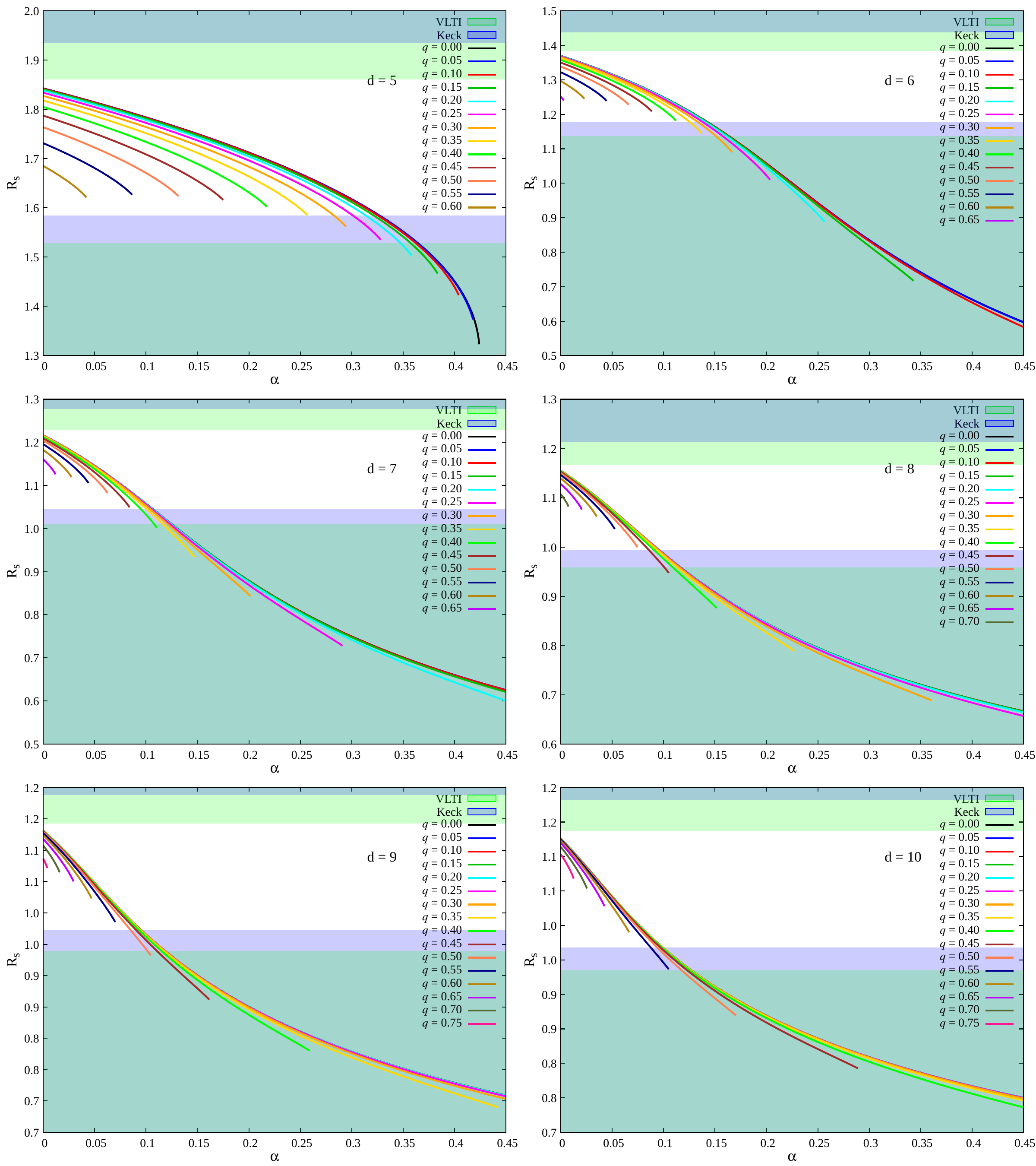}
\caption{The relevant parameters of the Bardeen-EGB black hole are constrained through the $d$-dimensional shadow constraint formula \eqref{eq:highdimrange}. The unfilled color region corresponds to the effective range $r_{\mathrm{sh}}$ of the shadow radius from the $\mathrm{Sgr~A}^{\star}$ data. The parameters $M=1$ and ${G}=1$ are selected.}
\label{FIG4}
\end{figure}

\begin{figure}[htbp]
\centering
\includegraphics[width=1\textwidth]{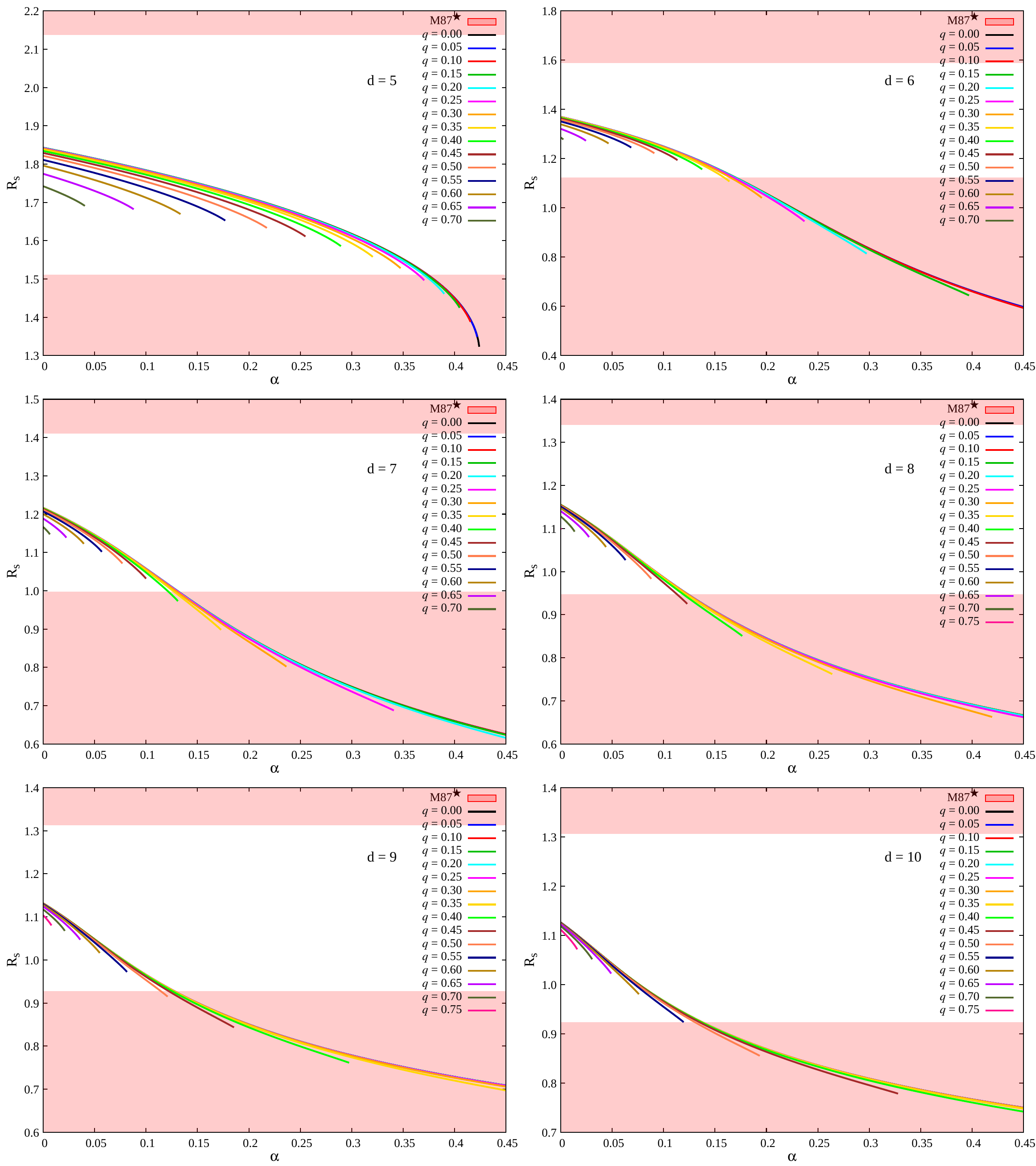}
\caption{The relevant parameters of the Hayward-EGB black hole are constrained through the $d$-dimensional shadow constraint formula \eqref{eq:highdimrange}. The unfilled color region corresponds to the effective range $r_{\mathrm{sh}}$ of the shadow radius from the $\mathrm{M87}^{\star}$ data. The parameters $M=1$ and ${G}=1$ are selected.}
\label{FIG5}
\end{figure}

\begin{figure}[htbp]
\centering
\includegraphics[width=1\textwidth]{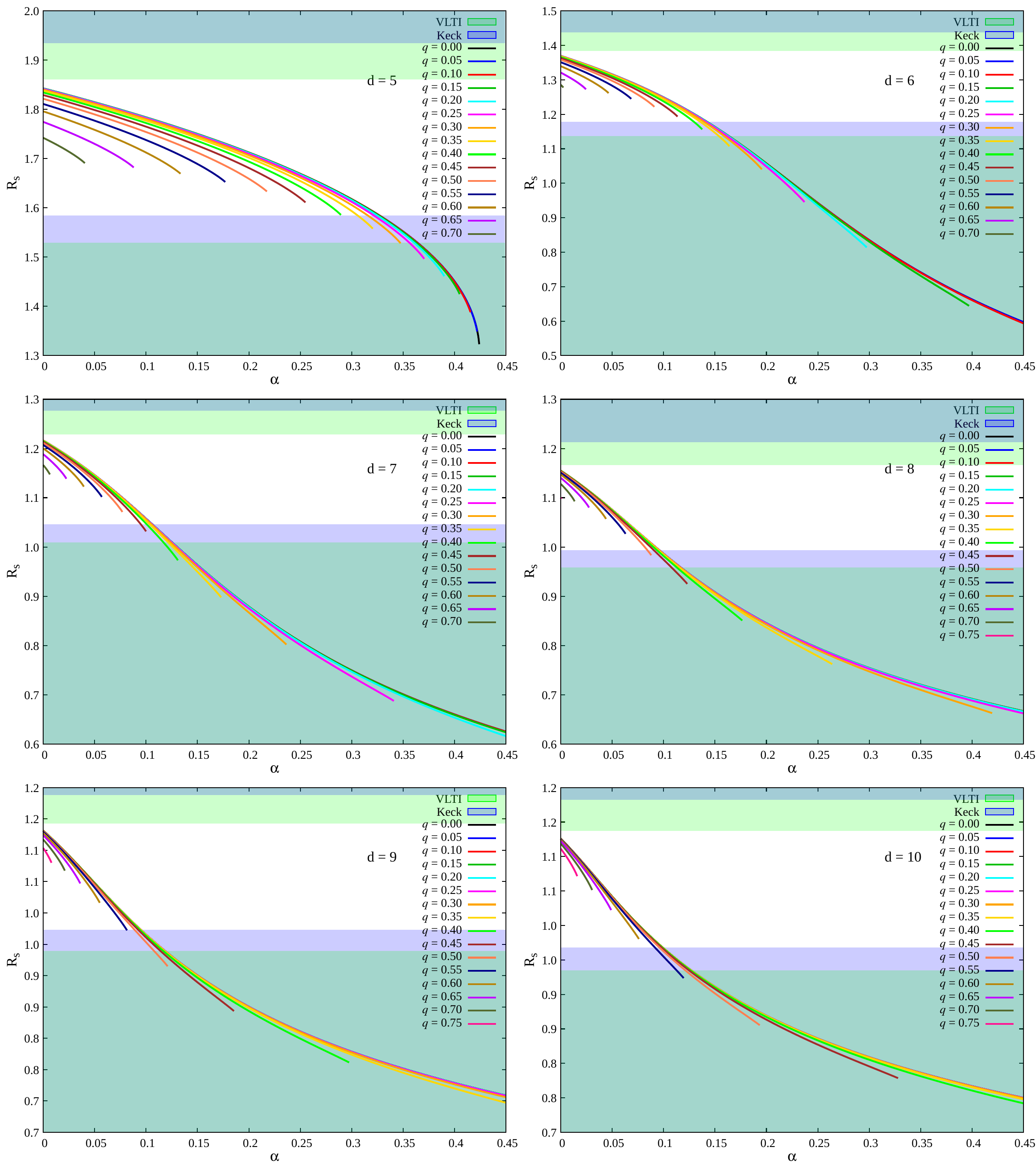}
\caption{The relevant parameters of the Hayward-EGB black hole are constrained through the $d$-dimensional shadow constraint formula \eqref{eq:highdimrange}. The unfilled color region corresponds to the effective range $r_{\mathrm{sh}}$ of the shadow radius from the $\mathrm{Sgr~A}^{\star}$ data. The parameters $M=1$ and ${G}=1$ are selected.}
\label{FIG6}
\end{figure}

Secondly, we apply the high-dimensional shadow radius formula \eqref{eq:highdimrs} to analyze the Bardeen-EGB and Hayward-EGB black hole solutions individually, and present the numerical results for $R_{s}$ within the admissible parameter space $(\alpha, q)$.
Subsequently, we employ the $d$-dimensional shadow constraint formula \eqref{eq:highdimrange} to provide constraints on the value of $R_{s}$, based on the observational data from the $\mathrm{M87}^{\star}$ and $\mathrm{Sgr~A}^{\star}$ independently.
All the above results will be shown in Figures \ref{FIG3}, \ref{FIG4}, \ref{FIG5} and \ref{FIG6}.

The comparative analysis reveals that both black hole configurations exhibit analogous characteristics: in various dimensions, the effective interval of the coupling constant $\alpha$ undergoes monotonic contraction with increasing magnetic charge $q$.
Crucially, this effective interval will progressively shrink to within the admissible region $r_{\mathrm{sh}}$ constrained by the $d$-dimensional shadow radius.
Since the value of the shadow radius $R_{s}$ decreases monotonically with the increase of the parameter $\alpha$, and $q=0$ (i.e., Eq. \eqref{eq:egbmetric}) corresponds to the maximum effective interval $(0, \alpha_{\text{max}})$ of the parameter $\alpha$, we will provide specific numerical results regarding the constraints of $\mathrm{M87}^{\star}$ and $\mathrm{Sgr~A}^{\star}$ (Keck) on the maximum interval $(0, \alpha_{\text{max}})$, as shown in Table \ref{tb2:alpharange}.

\begin{table}[tbh]\centering
\caption{The permissible range of the parameter $\alpha$ in the $d$-dimensional EGB black hole space-time is precisely determined by applying the shadow radius constraint. The parameters $q=0$, $M=1$ and ${G}=1$ are selected.
\vspace{0.2cm}} \label{tb2:alpharange}
\begin{tabular*}{10cm}{*{3}{c @{\extracolsep\fill}}}
\hline
$d$ & $\mathrm{M87}^{\star}$ & $\mathrm{Sgr~A}^{\star}\;\;(\mathrm{Keck})$ \\
\hline
5 & $0 < \alpha \lesssim 0.373956$ & $0 < \alpha \lesssim 0.326954$ \\
6 & $0 < \alpha \lesssim 0.170079$ & $0 < \alpha \lesssim 0.142480$ \\
7 & $0 < \alpha \lesssim 0.131696$ & $0 < \alpha \lesssim 0.106094$ \\
8 & $0 < \alpha \lesssim 0.123752$ & $0 < \alpha \lesssim 0.0959430$ \\
9 & $0 < \alpha \lesssim 0.127387$ & $0 < \alpha \lesssim 0.0950473$ \\
10 & $0 < \alpha \lesssim 0.137810$ & $0 < \alpha \lesssim 0.0989279$ \\
\hline
\end{tabular*}
\end{table}

\section{Conclusions}
In this study, we first establish a novel framework for constraining $d$-dimensional black hole solutions through shadow observation analysis.
We then provide a comprehensive constraint formula that governs higher-dimensional shadow radius by employing the Schwarzschild-Tangherlini metric modeling, ensuring that the constraints on the black hole shadow radius tend toward standardization.
Subsequently, we implement the framework by selecting two representative regular EGB black hole solutions as case studies, conducting detailed individual analyses through rigorous application of the $d$-dimensional shadow constraint formula.
Ultimately, we established the reasonable constraint ranges of the characteristic parameters in the two black hole metrics, demonstrating that our systematically developed framework exhibits universal applicability to broad classes of higher-dimensional black hole solutions.

\section*{Acknowledgements}
The author acknowledges all the individuals, teams, and institutions that contributed to the development of this paper.

\section*{Data Availability Statement} 
All relevant data are within the paper.

\bibliographystyle{plain}
\bibliography{D-SHADOW}

\begin{thebibliography}{10}

\bibitem{EventHorizonTelescope:2019dse}
Kazunori Akiyama et~al.
\newblock {First M87 Event Horizon Telescope Results. I. The Shadow of the
  Supermassive Black Hole}.
\newblock {\em Astrophys. J. Lett.}, 875:L1, 2019.

\bibitem{EventHorizonTelescope:2019ggy}
Kazunori Akiyama et~al.
\newblock {First M87 Event Horizon Telescope Results. VI. The Shadow and Mass
  of the Central Black Hole}.
\newblock {\em Astrophys. J. Lett.}, 875(1):L6, 2019.

\bibitem{EventHorizonTelescope:2021bee}
Kazunori Akiyama et~al.
\newblock {First M87 Event Horizon Telescope Results. VII. Polarization of the
  Ring}.
\newblock {\em Astrophys. J. Lett.}, 910(1):L12, 2021.

\bibitem{EventHorizonTelescope:2022wkp}
Kazunori Akiyama et~al.
\newblock {First Sagittarius A* Event Horizon Telescope Results. I. The Shadow
  of the Supermassive Black Hole in the Center of the Milky Way}.
\newblock {\em Astrophys. J. Lett.}, 930(2):L12, 2022.

\bibitem{EventHorizonTelescope:2022xqj}
Kazunori Akiyama et~al.
\newblock {First Sagittarius A* Event Horizon Telescope Results. VI. Testing
  the Black Hole Metric}.
\newblock {\em Astrophys. J. Lett.}, 930(2):L17, 2022.

\bibitem{EventHorizonTelescope:2024rju}
Kazunori Akiyama et~al.
\newblock {First Sagittarius A* Event Horizon Telescope Results. VIII. Physical
  Interpretation of the Polarized Ring}.
\newblock {\em Astrophys. J. Lett.}, 964(2):L26, 2024.

\bibitem{Cai:2001dz}
Rong-Gen Cai.
\newblock {Gauss-Bonnet black holes in AdS spaces}.
\newblock {\em Phys. Rev. D}, 65:084014, 2002.

\bibitem{Cho:2002hq}
Y.~M. Cho and Ishwaree~P. Neupane.
\newblock {Anti-de Sitter black holes, thermal phase transition and holography
  in higher curvature gravity}.
\newblock {\em Phys. Rev. D}, 66:024044, 2002.

\bibitem{Falcke:1999pj}
Heino Falcke, Fulvio Melia, and Eric Agol.
\newblock {Viewing the shadow of the black hole at the galactic center}.
\newblock {\em Astrophys. J. Lett.}, 528:L13, 2000.

\bibitem{Ghosh:2020tgy}
Sushant~G. Ghosh, Arun Kumar, and Dharm~Veer Singh.
\newblock {Anti-de Sitter Hayward black holes in
  Einstein\textendash{}Gauss\textendash{}Bonnet gravity}.
\newblock {\em Phys. Dark Univ.}, 30:100660, 2020.

\bibitem{Khodadi:2022pqh}
Mohsen Khodadi and Gaetano Lambiase.
\newblock {Probing Lorentz symmetry violation using the first image of
  Sagittarius A*: Constraints on standard-model extension coefficients}.
\newblock {\em Phys. Rev. D}, 106(10):104050, 2022.

\bibitem{EventHorizonTelescope:2021dqv}
Prashant Kocherlakota et~al.
\newblock {Constraints on black-hole charges with the 2017 EHT observations of
  M87*}.
\newblock {\em Phys. Rev. D}, 103(10):104047, 2021.

\bibitem{Kumar:2018vsm}
Arun Kumar, Dharm Veer~Singh, and Sushant~G. Ghosh.
\newblock {$D$-dimensional Bardeen-AdS black holes in Einstein-Gauss-Bonnet
  theory}.
\newblock {\em Eur. Phys. J. C}, 79(3):275, 2019.

\bibitem{Lu:2023bbn}
Ru-Sen Lu et~al.
\newblock {A ring-like accretion structure in M87 connecting its black hole and
  jet}.
\newblock {\em Nature}, 616(7958):686--690, 2023.

\bibitem{Medeiros:2023pns}
Lia Medeiros, Dimitrios Psaltis, Tod~R. Lauer, and Feryal Ozel.
\newblock {The Image of the M87 Black Hole Reconstructed with PRIMO}.
\newblock {\em Astrophys. J. Lett.}, 947(1):L7, 2023.

\bibitem{Nozari:2023flq}
Kourosh Nozari and Sara Saghafi.
\newblock {Asymptotically locally flat and AdS higher-dimensional black holes
  of Einstein\textendash{}Horndeski\textendash{}Maxwell gravity in the light of
  EHT observations: shadow behavior and deflection angle}.
\newblock {\em Eur. Phys. J. C}, 83(7):588, 2023.

\bibitem{Nozari:2024jiz}
Kourosh Nozari, Sara Saghafi, and Ali Mohammadpour.
\newblock {Higher-dimensional MOG dark compact object: shadow behaviour in the
  light of EHT observations}.
\newblock {\em Eur. Phys. J. C}, 84(8):778, 2024.

\bibitem{Pantig:2022qak}
Reggie~C. Pantig, Ali \"Ovg\"un, and Durmu\c{s} Demir.
\newblock {Testing symmergent gravity through the shadow image and weak field
  photon deflection by a rotating black hole using the M87$^*$ and Sgr. $\hbox
  {A}^*$ results}.
\newblock {\em Eur. Phys. J. C}, 83(3):250, 2023.

\bibitem{EventHorizonTelescope:2020qrl}
Dimitrios Psaltis et~al.
\newblock {Gravitational Test Beyond the First Post-Newtonian Order with the
  Shadow of the M87 Black Hole}.
\newblock {\em Phys. Rev. Lett.}, 125(14):141104, 2020.

\bibitem{Uniyal:2022vdu}
Akhil Uniyal, Reggie~C. Pantig, and Ali \"Ovg\"un.
\newblock {Probing a non-linear electrodynamics black hole with thin accretion
  disk, shadow, and deflection angle with M87* and Sgr A* from EHT}.
\newblock {\em Phys. Dark Univ.}, 40:101178, 2023.

\bibitem{Vagnozzi:2022moj}
Sunny Vagnozzi et~al.
\newblock {Horizon-scale tests of gravity theories and fundamental physics from
  the Event Horizon Telescope image of Sagittarius A}.
\newblock {\em Class. Quant. Grav.}, 40(16):165007, 2023.

\bibitem{Yan:2024rsx}
Zening Yan.
\newblock {Employing shadow radius to constrain extra dimensions in black
  string space-time with dark matter halo}.
\newblock 12 2024.

\bibitem{Yan:2023pxj}
Zening Yan, Xiaoji Zhang, Maoyuan Wan, and Chen Wu.
\newblock {Shadows and quasinormal modes of a charged non-commutative black
  hole by different methods}.
\newblock {\em Eur. Phys. J. Plus}, 138(5):377, 2023.

\end{thebibliography}

\end{document}